\documentclass[12pt,preprint]{aastex}
\pdfoutput=1

\slugcomment{Submitted to ApJ, July, 2013}

\shorttitle{Stability of the Outer Planets}
\shortauthors{Reyes-Ruiz et al.}

\begin{document}


\title{Stability of the Outer Planets in Multiresonant Configurations \\
        with a Self-gravitating Planetesimal Disk}


\author{M. Reyes-Ruiz\altaffilmark{1}, 
H. Aceves\altaffilmark{1}, and 
C. E. Chavez\altaffilmark{2}
}
\altaffiltext{1}{Universidad Nacional Aut\'onoma de M\'exico, Instituto de Astronom\'{\i}a, 
                      Apdo.Postal 106, Ensenada, B.C. 22860 M\'exico}
\altaffiltext{2}{Facultad de Ingenier\'{\i}a Mec\'anica y El\'ectrica, Universidad Aut\'onoma 
                      de Nuevo Le\'on, Monterrey, Nuevo Le\'on, 66451, M\'exico}


\email{maurey@astro.unam.mx}

\begin{abstract}
We study the effect of a massive planetesimal disk on the dynamical stability of the outer 
planets assuming, as has been suggested recently, that these were initially locked in a 
compact and multiresonant configuration as a result of gas-driven migration in a protoplanetary 
disk. The gravitational interaction among {\em all} bodies in our simulations is included 
self-consistently using the {\tt Mercury6.5} code. Several initial multiresonant configurations 
and planetesimal disk models are considered. Under such conditions a strong dynamical 
instability, manifested as a rapid giant planet migration and planetesimal disk dispersal, develops 
on a timescale of less than 40 Myr in most cases. Dynamical disk heating due to the gravitational interactions 
among planetesimals leads to more frequent interactions between the planetesimals and the ice giants 
Uranus and Neptune, 
in comparison to models in which planetesitmal-planetesimal interactions are neglected. On 
account of the rapid evolution of the multiresonant configurations obtained with fully 
self-consistent simulations, our results are inconsistent with the dynamical instability origin 
of the Late Heavy Bombardment as currently considered by the Nice model for the Solar System. 
\end{abstract}


\section{Introduction}

The formation of the Solar System is generally believed to involve, during
its last phases, a period of significant radial migration of the giant planets due
to their interaction with a massive, remnant planetesimal disk located outward 
of the outermost planet (for a review see Levison et al. 2007). 
This planetesimal-driven migration involves the 
inward migration of Jupiter, the outward migration of Saturn, Uranus and 
Neptune, and the dispersal of most of the planetesimals in the disc into orbits 
at much greater distances from the Sun. 

The current scenario has evolved from the original suggestion of 
Fernandez \& Ip (1984), then developed by Malhotra (1993),  in which the
outer planets migrate as they exchange angular momentum efficiently with 
the planetesimals in the outer Solar System while the planetesimals are dispersed to 
the Oort cloud and beyond. 
Early numerical simulations of this process (Hahn \& Malhotra 1999, 
Gomes et al. 2004) indicated that Jupiter migrates inward typically 
by a fraction of an AU, while Saturn, Uranus and Neptune migrate outwards 
changing their initial semimajor axis by approximately 10, 50 and 100\%, respectively.

In a series of papers by Gomes et al. (2005), Tsiganis et al. (2005) and Morbidelli et al. (2005), 
a variant of the migration scenario proposed by Malhotra (1993, 1995) was presented. 
As part of what is now known as the Nice model for the early evolution of the Solar System,
an initially very compact configuration of the outer Solar System was proposed as the condition
at the time the primordial gaseous nebula was dispersed. A distinctive feature of the initial 
configuration originally proposed in the Nice model was the fact that it is unstable 
on a timescale of several hundred megayears, with Saturn crossing the 2:1 mean motion 
resonance (MMR) with Jupiter, and leading to a major rearrangement of the outer Solar System.
  
In recent years, the Nice model has undergone major revisions. The model now has
incorporated the idea that in the initial orbits of the outer planets are in a multi-resonant 
configuration, as expected following an {\it apriori} gas-driven migration phase 
(Masset \& Snellgrove 2006, Morbidelli et al. 2007 and  Batygin \& Brown 2010, among others). 
Such configurations are known to be stable 
if only the gravitational interaction among the planets is considered 
(Batygin \& Brown 2010). Secondly, as argued by Brasser et al. (2007), 
the stability of the inner solar system during the period of outer planet 
migration seems to require an abrupt change in Jupiter's orbit 
(see also Minton \& Malhotra 2009). This change can in principle result from a close encounter 
of Jupiter with an ice giant that may eventually be ejected from the solar system,
leading to the so-called ``jumping Jupiter'' revision of the Nice model 
(Morbidelli et al. 2009b, 2010, Brasser et al. 2009,  Walsh \& Morbidelli 2011, 
Agnor \& Lin 2012). Finally, by using an approximate treatment for the 
self-gravitational interaction in the planetesimal disk (an effect not previously 
considered), Levison et al. (2011) suggested that, starting from a stable  
multi-resonant configuration, self-gravity instabilities slowly develop and 
can lead to the rapid migration of Neptune into the planetesimal disk, on a 
timescale consistent with the occurrence of the Late Heavy Bombardment (LHB) 
several hundred megayears after gas disk dispersal.  

In this paper we present a series of $N$-body simulations that probably resemble the 
dynamical evolution the outer Solar System had after the primordial gaseous nebula 
was dispersed, 
in order to study in some detail the 
stability of the system taking into account {\em self-consistently} the 
gravitational interactions among the planetesimals. To our knowledge, 
with the exception of the work by Levison et al. (2011), previous 
studies of planetesimal driven migration have considered these small bodies to 
interact gravitationally only with the planets but not among themselves. This  
approximation is only justified on account of the great computational expense of 
the direct $N$-body simulations required to model this process. In this paper, 
we present results of the first numerical simulations to study the effect of 
a self-gravitating disk on the process of planetesimal driven migration in 
a fully self-consistent manner.  

The paper is organized as follows. In Section~\ref{sec:model} we present the details of 
our model, we describe the methodology used and the 
cases studied. In $\S$\ref{sec:results} we present the results of a series 
of simulations with different model parameters. A review of some of the
assumptions made in our study and some of the implications for the 
scenarios of planetary system evolution are discussed in $\S$\ref{sec:discussion}. 
Finally, in Section~\ref{sec:conclusions} we summarize our main results and 
present our conclusions.


\section{Model Description}
\label{sec:model}

Models of the process of planetesimal driven migration generally assume
that the process begins in earnest once the gaseous component of the protoplanetary 
disk is dispersed somehow, at least to the point that dynamical effects of the gas 
can be neglected. In this study we adopt this scenario in order to make a direct 
comparison to results from other authors. We note, however, that preliminary 
results of an ongoing investigation by our group, suggest that the 
remnant gaseous component of protoplanetary disks during the so-called transition 
disk phase, can have important consequences on the evolution of the planetesimal 
population and the outer planets. 

We consider the gravitational force of the Sun, the 4 outer planets with their 
present day mass, and a massive disk of much smaller bodies representing the 
remnant planetesimal disk exterior to the planetary region. A novel feature of 
our simulations, in comparison to previous studies of planetesimal-driven migration, 
is the inclusion of the gravitational force among the planetesimals themselves in 
a fully self-consistent manner.

\subsection{Planetary configuration}

In accordance to the latest versions of the Nice model, based on the results of 
Masset \& Snellgrove (2006) and other authors (e.g. Morbidelli et al. 2007, 
Nesvorny \& Brown 2010 and Zhang \& Zhou 2010), we assume that initially  
the outer planets were locked in mean motion resonance (MMR) with each other. 
Specifically, we adopt some of the multiresonant configurations described in 
Table~\ref{tbl-1} of Batygin \& Brown (2010) as a frequent outcome of their study
of resonant trapping in gaseous disks. In our notation, a 3J:2S, 4S:3U, 4U:3N 
configuration is one in which Jupiter and Saturn are in a 3:2 MMR (i.e. Jupiter
makes 3 revolutions around the Sun while Saturn orbits twice), Saturn and Uranus 
are in a 4:3 MMR and Uranus and Neptune are also in a 4:3 MMR.

\begin{figure}
\includegraphics[width=\columnwidth]{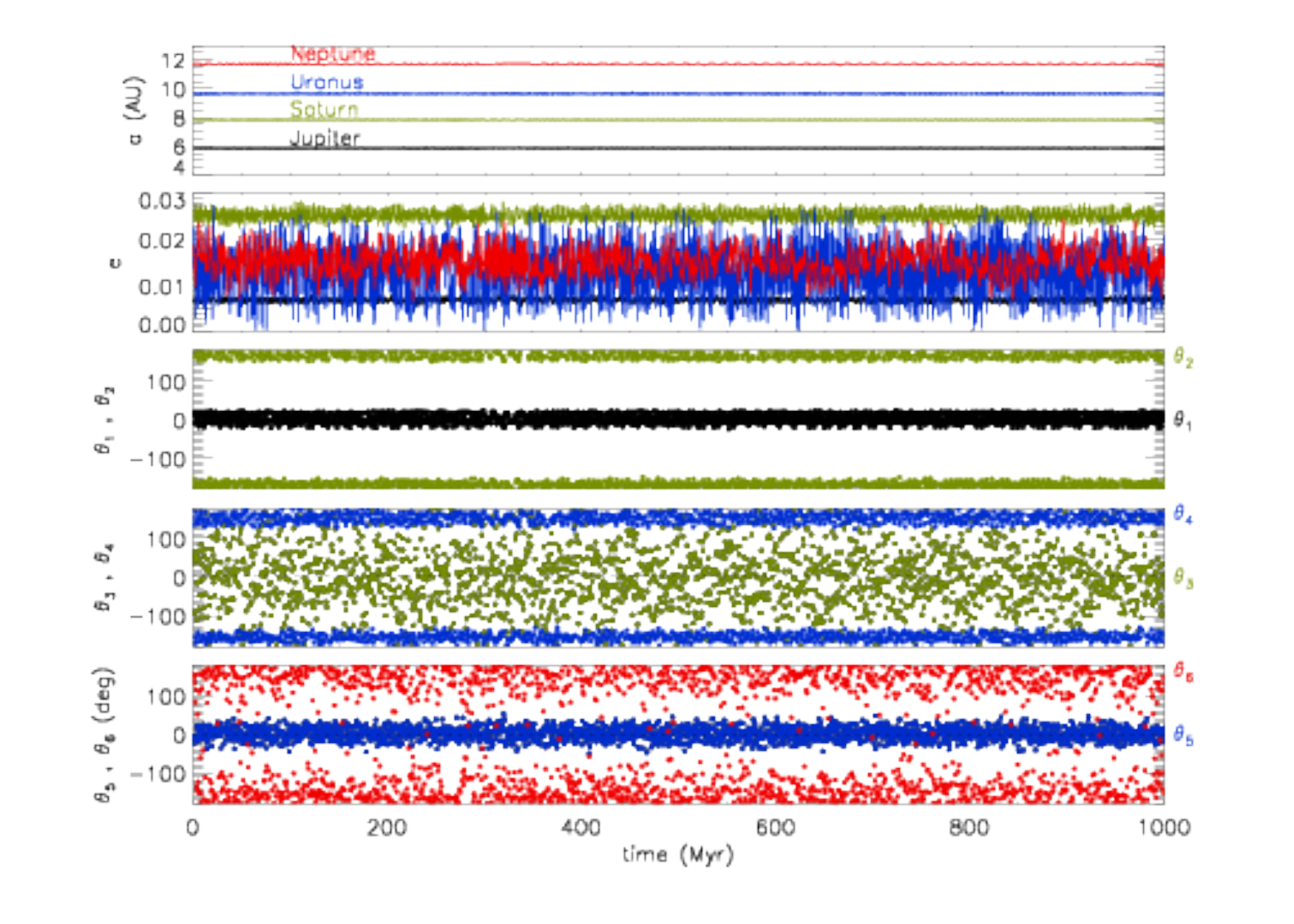}
\caption{Evolution of the orbital elements, $a$ and $e$, and of the resonant angles given 
by Eqn. (\ref{res_arg}) for the planets in the 3J:3S,4S:3U,4U:3N multiresonant configuration 
used in our simulations, without a planetesimal disk. The orbits are coplanar throughout the 
evolution. 
\label{fig1}}
\end{figure}

The precise value of the main orbital parameters used for these initial conditions are
shown in Table \ref{tbl-2} (more details can be 
obtained by mailing the corresponding author). These configurations are dynamically 
stable (without a planetesimal disk) over gigayear timescales. As can be seen 
in Figure~\ref{fig1},  which shows the evolution of the semimajor axis, $a$, 
and eccentricity, $e$, of each 
of the planets considered in case A of our simulations. The behavior of the other multiresonant 
planetary configurations we have considered without the planetesimal is very similar.  
For reference, we include in Table \ref{tbl-2} the average value of this parameters
corresponding to the  3J:2S, 3S:2U, 4U:3N 
multiresonant system considered by Levison et al. (2011).

\subsection{Planetesimal disk model}

Following Levison et al. (2011), who studied the effect of the planetesimal disk  
self-gravity on the equilibrium of the system for the first time in an approximate 
manner, we assume that 
initially the disk extends from a radius located outwards of the outermost planet, 
$R_{\rm disk}$, to an outer radius approximately at the distance of Neptune's current 
orbital semimajor axis, $R_{\rm out}$. The inner disk radius is taken as a parameter in our model 
with several values considered as given in Table~1. Note that the Hill radius for Neptune 
at the corresponding distance in these compact configurations is approximately 0.31 AU,
so that all values of $R_{\rm disk}$ considered are more than 3 Hill radii from the outermost
planet; i.e. outside its so-called feeding zone. The outer radius for all cases considered is 
taken at 31.5 AU, in concordance with the study of Levison et al. (2011).

On account of the computation time required for these simulations, the disk is restricted 
to be made of $N_{\rm p} = 2000$ particles for most of our calculations. Although our present 
computational capabilities do not permit considering a much greater number of particles,
in the Discussion section we address the probable implications of this choice briefly. The mass of
each particle is then $m_{\rm p} = M_{\rm disk}/N_{\rm p}$, where $M_{\rm disk}$ 
is the total disk mass. For the disk mass values adopted  in this study (Table 1), the 
planetesimal mass ranges from $\sim$ 0.01 to 0.05 $M_{\rm E}$ or between 1 and 4 Lunar 
masses approximately; where $M_{\rm E}$ is the mass of the Earth. Although this 
mass is still a factor of a few greater than the average mass of the dwarf 
planets, which can probably be considered the largest planetesimals, including a greater number 
of planetesimals, with a self consistent treatment of gravitational forces, is beyond the limit of 
our present computational capabilities.

The mass of particles we have considered is comparable 
to that in the studies of Hahn \& Malhotra (1999), which is in the range 0.01 and 0.2 $M_{\rm E}$, 
and also of Gomes et al. (2004), who consider planetesimals of approximately 0.02 $M_{\rm E}$. 
In both cases however, planetesimal-planetesimal interactions were not considered. The planetesimal 
mass we have considered is also comparable to the mass considered for gravitational perturbers 
in the study of Levison et al. (2011). 

The mass surface density in the planetesimal disk, $\Sigma$, is assumed to be initially 
axisymmetric and to have a power law dependence on the radial distance from the 
Sun, $\Sigma = \Sigma_0 R^{-1}$, following Levison et al. (2011) and consistent 
with minimum mass solar nebula models. The value of $\Sigma_0$ is adopted to 
reproduce the total disk mass in a specific model. The planetesimal disk is assumed to be cold, 
with eccentricities, $e$, and inclinations, $i$ (in radians) chosen randomly from 
uniform number distributions in the range $[0, 10^{-3}]$. The rest of the orbital elements, 
$\Omega$, $\omega$ and $M$, are also chosen from a uniform random distribution 
within the range $[0,360^{\rm o}]$.

\subsection{Numerical details}

We use the {\tt Mercury6.5} code, developed by Chambers (1999), to study the dynamics 
of the four outer planets and the planetesimal disk. It uses a fixed time-step for the 
symplectic integrator coupled with a Bulirsch-Stoer (B-S) scheme, which is used to follow close encounters among 
bodies: we set the error tolerance of the B-S integrator to 10$^{-12}$.  The B-S scheme is 
used whenever the distance between two bodies is less than 3 Hill radii of the largest body.    

Our simulations were in general carried out for up to 70 Myr. 
The relatively short timescale we have considered is related to the main 
purpose of our study,  which is to assess the stability of these systems and not 
to determine the long-term final planetary configuration. 
The time-step used in our simulations is 186.4 days, as commonly adopted 
in outer solar system migration simulations (e.g. Hahn \& Malhotra 1999). 
The simulations presented in this paper were carried out in a Linux PC, with
OpenSuse 11.4 as operating system and a {\tt gfortran}-2.4 compiler.

\section{Results}\label{sec:results}

We have carried out a series of simulations aimed to understand the conditions under 
which planetesimal-driven migration, starting from multiresonant planetary configurations, 
becomes unstable. To this end, several initial conditions for the planets and for the 
planetesimal disk, were considered (see Table~1). In the following we present our 
results by considering the effect of varying the disk mass and inner radius for 
each of the multiresonant initial conditions studied.  

For comparison, we have computed cases A, F and K (described in Table~1) with
the same disk properties considered by Levison et al.~(2011) as their fiducial disk model;
$M_{\rm disk} = 50$ M$_{\rm E}$ and $R_{\rm disk} = 18$ AU. However, the 
planetary configurations we consider (see Tables 1 and 2), differ slightly from 
the 3J:2S, 3S:2U, 4U:3N multiresonant system that these authors study. 
Levison et al.~(2011) take their initial planetary configuration from Morbidelli et 
al. (2007), and report and average semimajor axis for Jupiter approximately 10\% smaller 
than in our cases. As a consequence, the initial semi-major 
axis of the outer planet in the configuration considered by Levison et al.~(2011), is 
smaller in all but the 3J:2S,4S:3U,4U:3N configuration we consider. This configuration
is the most compact we studied and also the one that destabilizes at an earlier 
time for all cases.

\begin{figure}
\includegraphics[width=\columnwidth]{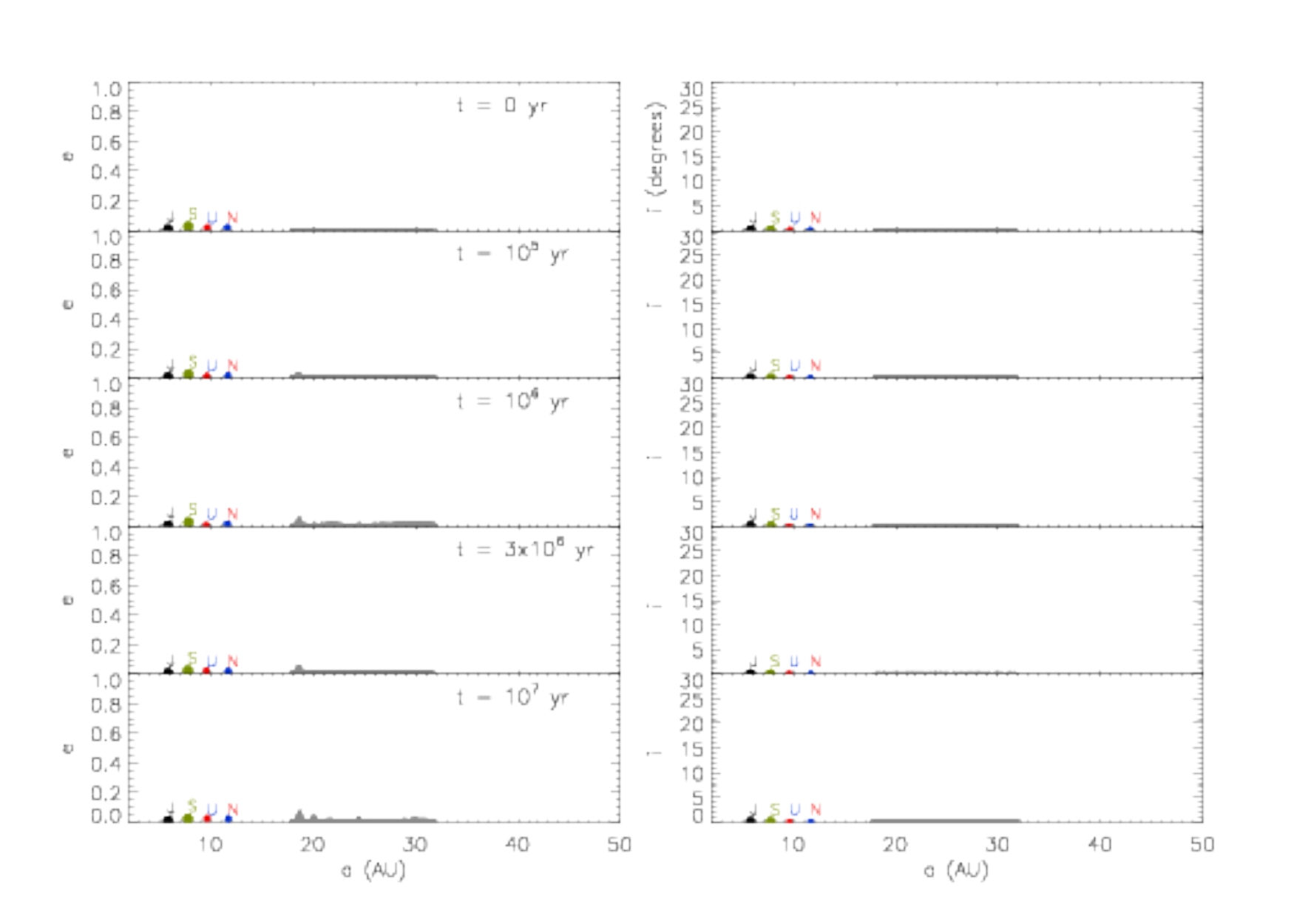}
\caption{Snapshots of the evolution of the orbital configuration of a system, 
$a$ {\it versus} $e$ (left column) and $a$ {\it versus} $i$ (right column), 
for a disk as in case A described in Table 1 but NOT considering the effect
of planetesimal-planetesimal interactions.  It illustrates a typical evolution of 
systems with Nice-like initial conditions which remain ``stable'' for 10 Myr.  
Small gray circles represent the planetesimals, black circles corresponding to 
the planets are labeled by their initial.  
\label{fig2}} 
\end{figure}

\begin{figure}
\includegraphics[width=\columnwidth]{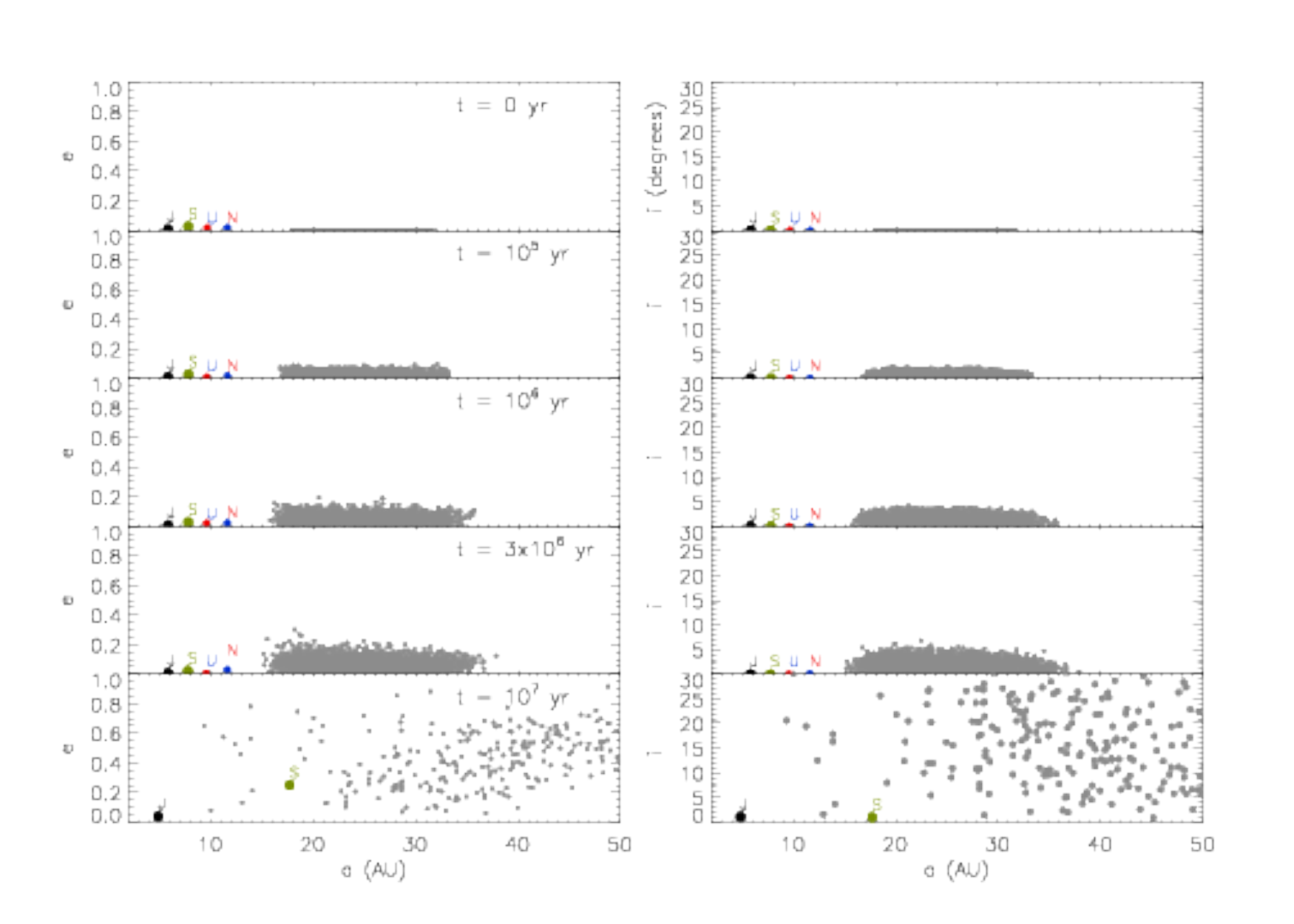}
\caption{Same as Figure \ref{fig2} but for a disk with planetesimal-planetesimal interactions
corresponding to case A described in Table 1.  It illustrates a typical unstable evolution of 
fully self-gravitating systems which become unstable on a short timescale.   
\label{fig3}} 
\end{figure}

The main effect of planetesimal-planetesimal interactions on the evolution of the system
can be seen comparing Figure \ref{fig2} and Figure \ref{fig3} which show the evolution of 
systems without and with self-gravity in the planetesimal disk, respectively. Both figures
shows several snapshots of the $a-e$ and $a-i$ phase space domain, of a system corresponding 
to the configuration of case A. Without the effect of self-gravity, Figure \ref{fig2}, 
the system is not unstable on the timescale considered, 10 Myr. The behavior of such system 
is consistent with the results of typical planetesimal-driven migration simulations starting 
from Nice model-like initial conditions.  If we include the gravitational
interaction among planetesimals, the system undergoes a dynamical instability in approximately 
5 Myr. Comparing the first and second row panels 
of Figure~\ref{fig3}, which illustrate the system's dynamical state initially and after 0.1 Myr respectively, 
we see that one of the main effects of the inter-planetesimal gravitational interactions is the heating 
of the disk, i.e. the increase in orbital eccentricities and inclinations also referred to as gravitational 
stirring. We discuss this process in more detail, and its dependence on the number of particles used 
in our simulation, in the following section. 

As can be seen in Figure \ref{fig3}, the disk remains in a quiescent state for a few Myr 
(for most cases considered) as the number of planetesimals dispersed into orbits capable 
of having close encounters with the outermost planets increases. The cumulative effect of these 
interactions, exchanging angular momentum with Neptune and Uranus mainly, leads to a growth of 
the orbital eccentricity for these planets until the multiresonant state among the giant planets 
is broken. The instability manifests as a very fast orbital rearrangement 
of all planets. The outward migration of Uranus and Neptune in turn leads 
to a greater number of planetesimal-planet interactions and even faster migration into the disk
until it is mostly dispersed. In several cases studied, the planetary system is strongly unstable with 
one or both ice giants ejected from the system, or even catastrophic as with one of them colliding 
with the giant planets as indicated by the numerical simulations.

\subsection{Effect of the disk mass}

In order to assess the effect of the planetesimal disk mass on the evolution of each of our initial 
multiresonant configurations, we have carried out simulations with $M_{\rm disk} = 25, 50$ 
and $100 M_\Earth$. In all cases, we consider the same inner and outer disk radii of the 
fiducial model of Levison et al. (2011). Disk mass is varied by changing the mass of each of 
the particles representing the planetesimals keeping the number of planetesimals constant.

\begin{figure}
\includegraphics[width=\columnwidth]{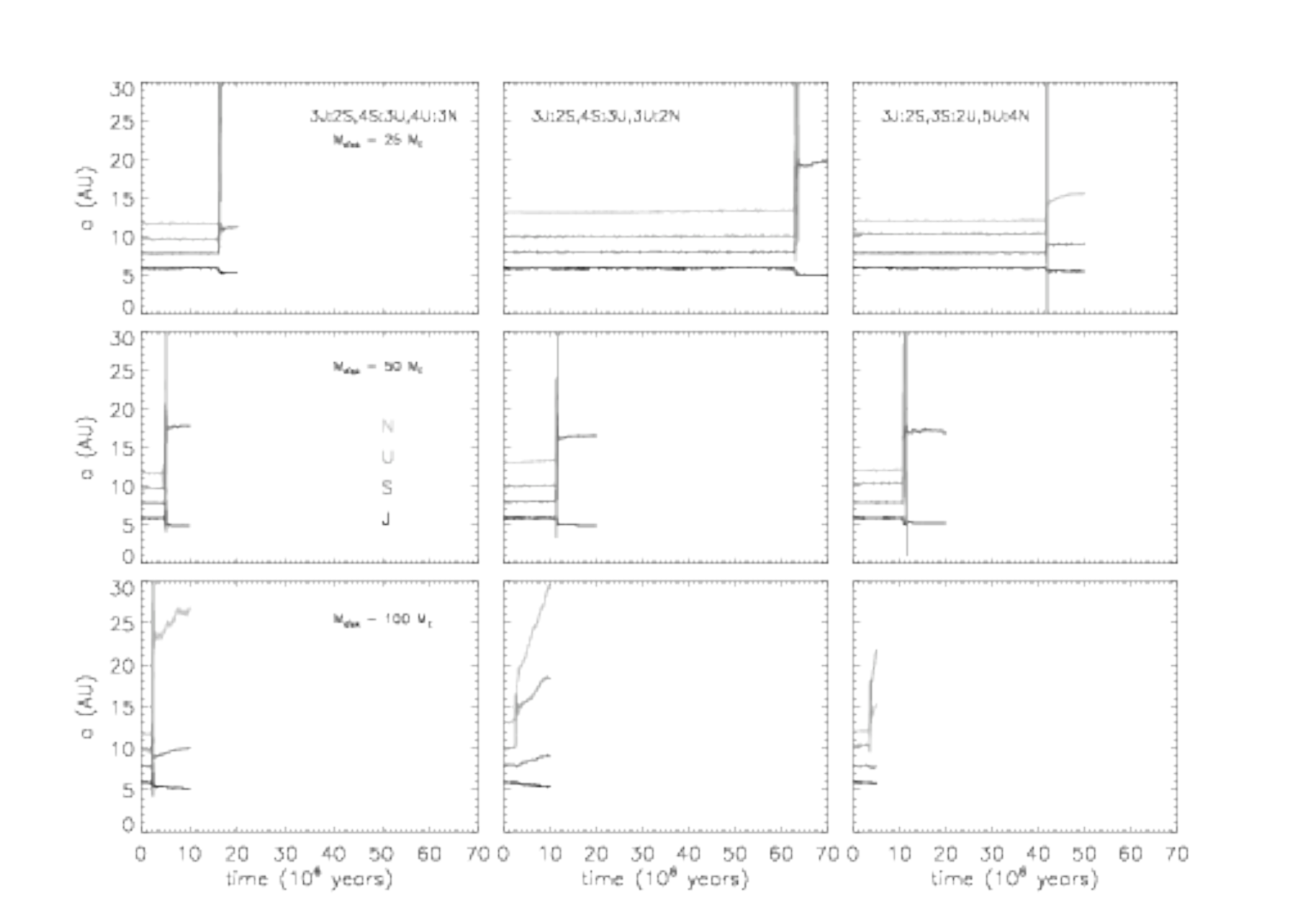}
\caption{Evolution of the planetary orbital semimajor axis for each of the 
             planetesimal disk masses considered. Columns correspond to different 
             multiresonant planetary configurations. The left column corresponds to  
		cases B, A and C, the middle column to cases G, F and H, and the 
		right-hand column illustrates the for cases L, K and M.
\label{a_vs_t_vs_Md}}
\end{figure}

\begin{figure}
\includegraphics[width=\columnwidth]{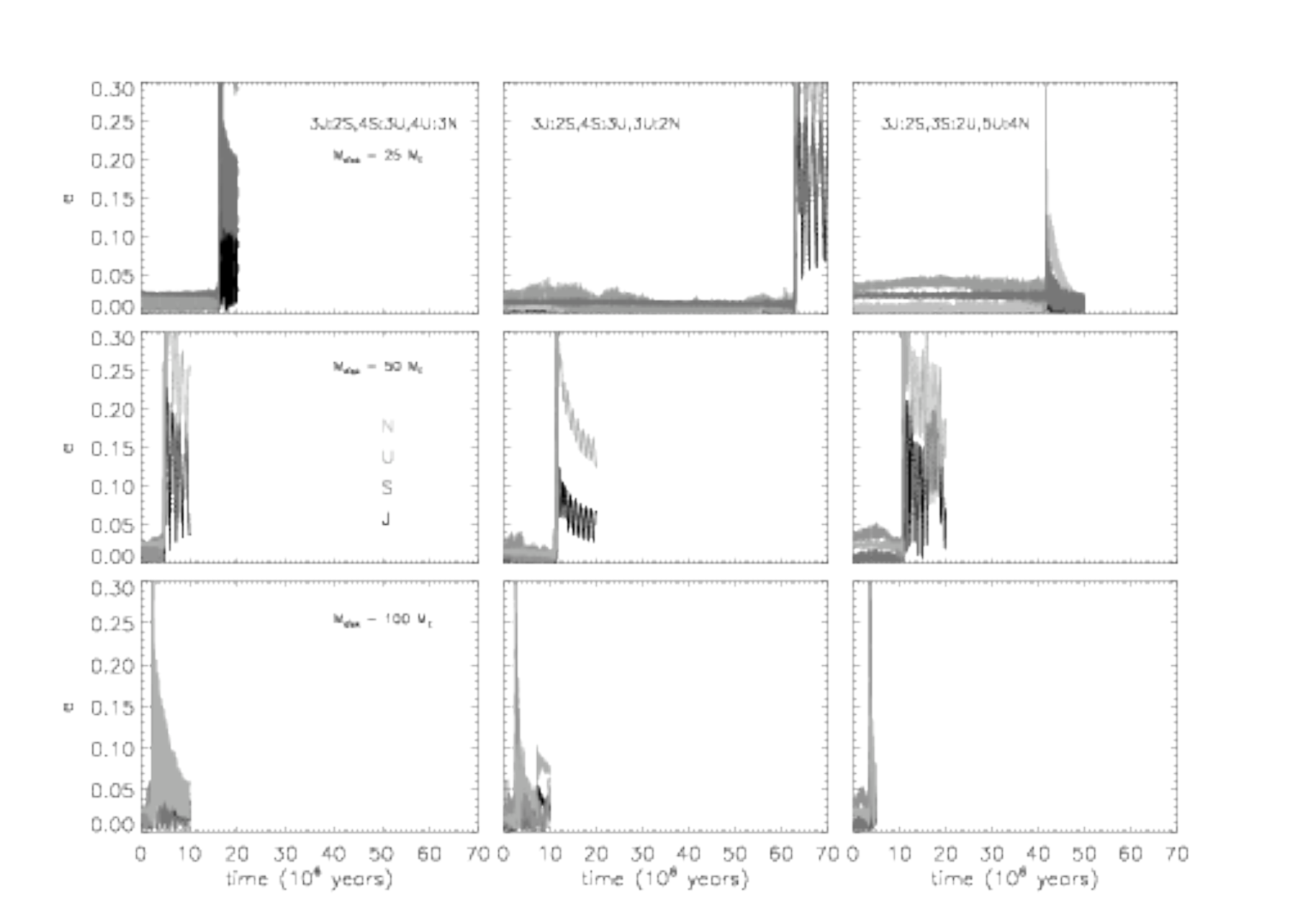}
\caption{Evolution of the planetary orbital eccentricity for each of
             the planetesimal disk masses considered. The left column corresponds to  
		cases B, A and C, the middle column to cases G, F and H, and the 
		right-hand column illustrates the for cases L, K and M.
\label{e_vs_t_vs_Md}}
\end{figure}

The time evolution of the planetary orbital parameters, $a$ and $e$, for all three multiresonant 
configurations we considered are shown in Figures~\ref{a_vs_t_vs_Md} and \ref{e_vs_t_vs_Md}, 
respectively. In these figures each row corresponds to a different planetary multiresonant  
initial condition. We find that, as expected, the instability arises earlier for more massive disks in 
which gravitational stirring increases the amount (and cumulative mass) of planetesimals interacting 
with the planets. 

With the exception of case G, all systems considered are unstable on a timescale shorter than 
42 Myr, as a result of the interaction with the self-gravitating planetesimal disk. Case G also becomes 
unstable but on a slightly greater timescale, less than 70 Myr. It is worth noting 
that case G corresponds to the combination of: (a) the less massive disk we have considered 
and (b) the planetary configuration in which Uranus and Neptune are most separated. 

\subsection{Effect of the disk inner radius}

\begin{figure}
\includegraphics[width=\columnwidth]{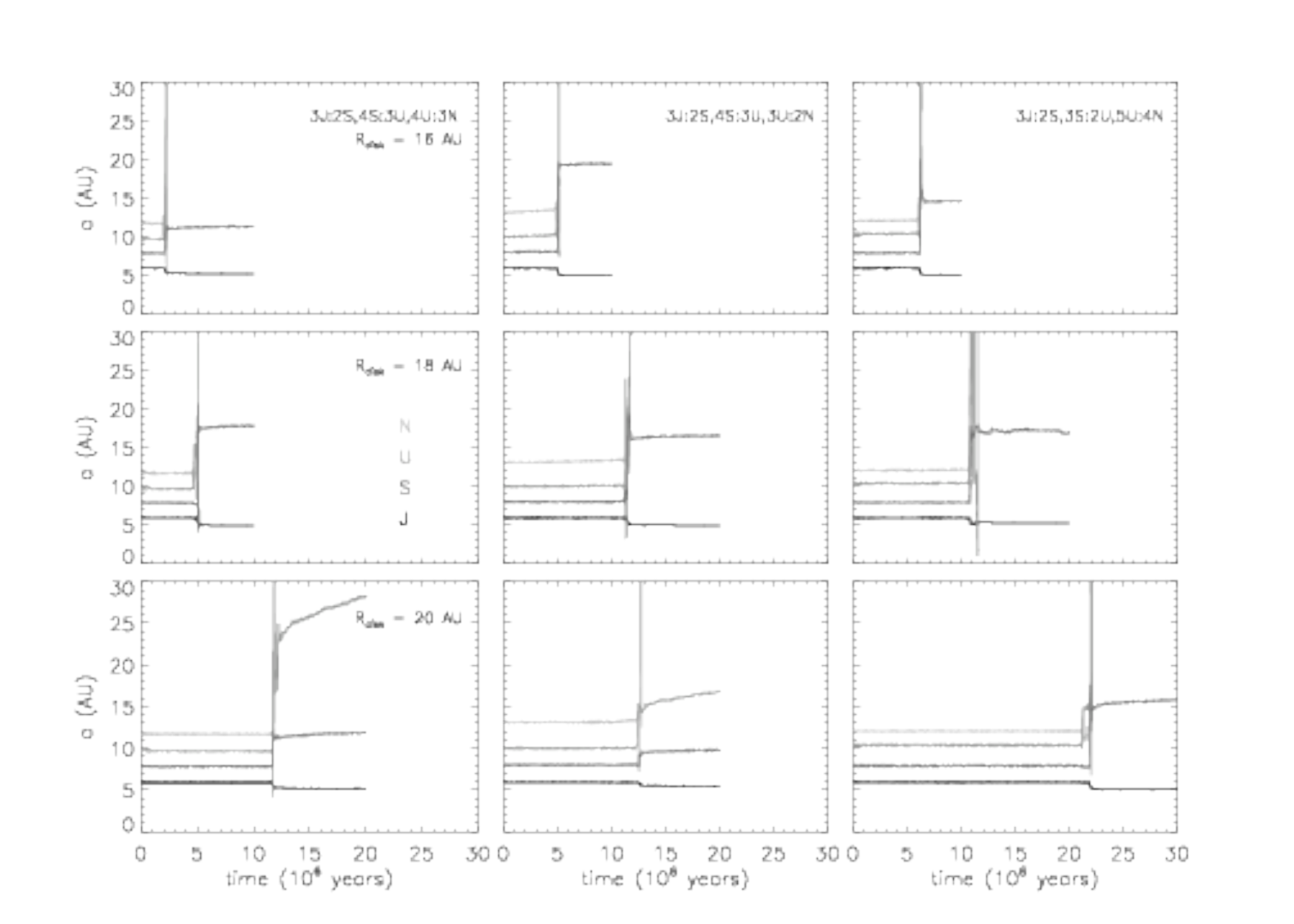}
\caption{Evolution of the planetary orbital semimajor axis for each of 
             the planetesimal disk inner radii considered. Columns correspond to different 
             multiresonant planetary configurations. The left column corresponds to 
             cases D, A and E, the middle column to cases I, F and J, and the right-hand
             column illustrates the for cases N, K and O.
\label{a_vs_t_vs_Rd}}
\end{figure}

\begin{figure}
\includegraphics[width=\columnwidth]{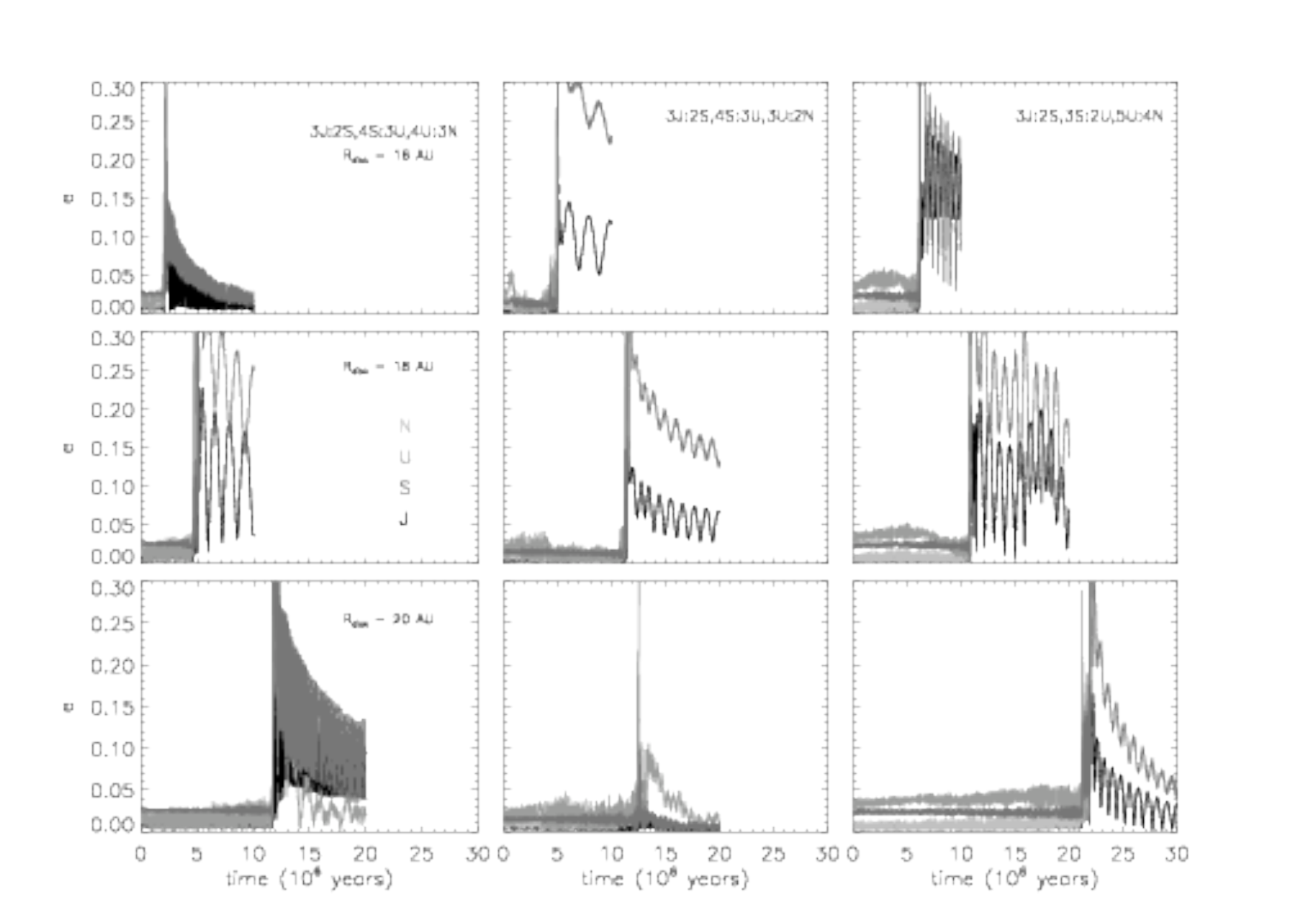}
\caption{Evolution of the planetary orbital eccentricity for each of the 
             planetesimal disk inner radii considered. The left column corresponds to 
             cases D, A and E, the middle column to cases I, F and J, and the right-hand
             column illustrates the for cases N, K and O. 
\label{e_vs_t_vs_Rd}}
\end{figure}

We have also studied the effect of the location of the inner radius  of 
the planetesimal disk, $R_{\rm disk}$, on the stability of these systems. To do so, 
we fix the disk mass at 
50 $M_{\Earth}$, following Levison et al.~(2011), and vary $R_{\rm disk}$ between 16 
and 20 AU. The results for all initial planetary configurations as a function of $R_{\rm disk}$ 
are displayed in Figures~\ref{a_vs_t_vs_Rd} and \ref{e_vs_t_vs_Rd}, which show 
the evolution of the semimajor axis, $a$, and eccentricity, $e$, respectively, for each of the 
planets in our simulations.

We find that the time for the onset of the instability depends strongly on the value 
of $R_{\rm disk}$. In going from an inner disk radius of 16 AU to 20 AU (a 25\% increment), 
the time for onset of the instability, $t_{\rm inst}$, increases by a factor of about 2-6.
This result is reminiscent of the results of Gomes et al. (2005) who find that
the initial location of the inner disk radius must be fine tuned to a specific value 
in order to reproduce the LHB timescale, if this phenomenon is to be related to a 
planetary instability. Once again, as in our simulations with different disk mass, 
all cases considered with a 50 $M_{\Earth}$ planetesimal disk, the outer Solar 
System becomes unstable within a time scale of 30 Myr. For comparison, we have 
carried out simulations of the three cases with different inner radius, for the 
3J:2S, 4S:3U, 4U:3N configuration and the same initial distribution of planetesimals, 
but without the effect of planetesimal-planetesimal interactions. In all cases, 
the system remains stable for a timescale an order of magnitude greater, or more, than 
that with selfgravitating planetesimal disks.

\subsection{Insight into the instability}

\begin{figure}
\includegraphics[width=\columnwidth]{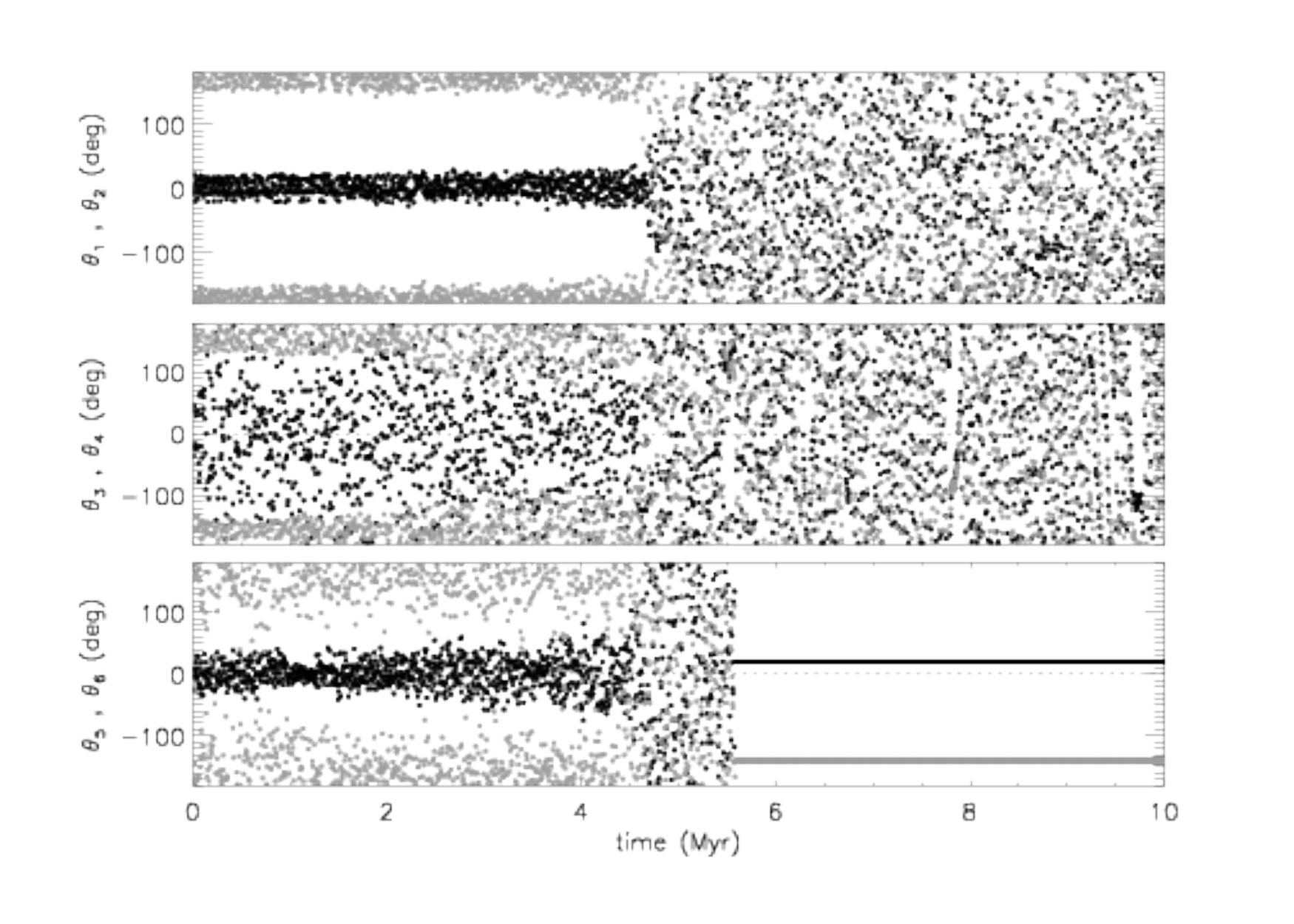}
\caption{Time evolution of the 2 body resonant arguments for each pair of planets
             as defined in Eqs.~(\ref{res_arg}) for case A. The top 
             panel corresponds to the 3:2 MMR resonance between Jupiter and Saturn, the 
             middle panel to the 4S:3U resonance and the bottom panel shows the 
             arguments of the 4U:3N MMR. In each panel, the 
             black dots denotes the first of the 2 arguments shown (e.g. $\theta_1$ in
             the top panel) and the gray dots show the second argument for that resonance 
             (e.g. $\theta_2$ in the top panel) .
\label{res_ang}}
\end{figure}

An additional feature of the evolution of the unstable systems, is the fact that the planets remain
locked in resonance until the instability arises. This is evident in Figure \ref{res_ang} which shows 
the mean motion resonant angles which, for the 3J:2S, 4S:3U, and 4U:3N initial planetary configuration, can 
be written as:
\begin{equation}
\theta_1 = 2 \lambda_{\rm J} - 3 \lambda_{\rm S} + \varpi_{\rm S} , 
\label{res_arg}
\end{equation}
$$
\theta_2 = 2 \lambda_{\rm J} - 3 \lambda_{\rm S} + \varpi_{\rm J} , 
$$
$$
\theta_3 = 3 \lambda_{\rm S} - 4 \lambda_{\rm U} + \varpi_{\rm U} ,
$$
$$
\theta_4 = 3 \lambda_{\rm S} - 4 \lambda_{\rm U} + \varpi_{\rm S} ,
$$
$$
\theta_5 = 3 \lambda_{\rm U} - 4 \lambda_{\rm N} + \varpi_{\rm N}  \ \ {\rm and} 
$$
$$
\theta_6 = 3 \lambda_{\rm U} - 4 \lambda_{\rm N} + \varpi_{\rm U} ,
$$
\noindent where $\lambda_i = M_i +\omega_i + \Omega_i$ is the mean longitude for the $i$-th planet, 
with $M_i$ being the mean anomaly, $\omega_i$ the argument of the planet's periapsis and 
$\Omega_i$ the longitude of the ascending node. Also, $\varpi_i$ is the longitude of the orbit's
periapsis.

The behavior seen in Figure \ref{res_ang} is qualitatively similar for all systems before the 
instability develops, namely:
(a) the $\theta_1$ and $\theta_2$ resonant arguments for Jupiter and Saturn librate 
about 0 and 180$^o$, respectively. (b) Either the $\theta_3$ or $\theta_4$ resonant 
arguments for Saturn and Uranus clearly librate while the other argument may circulate, 
and (c) at least one of the resonant arguments $\theta_5$ or $\theta_6$ for Uranus and 
Neptune librates.

\begin{figure}
\includegraphics[width=\columnwidth]{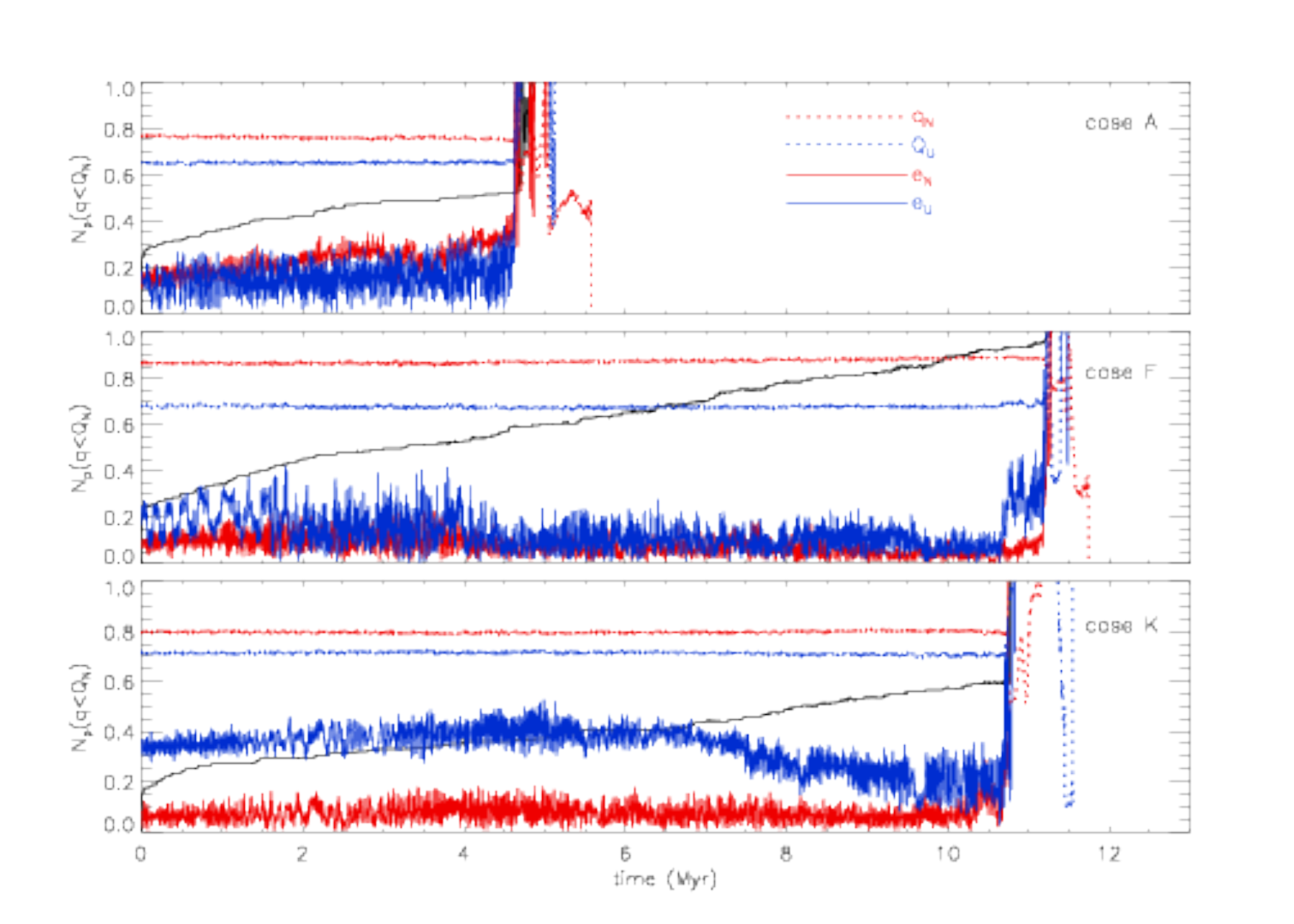}
\caption{Time evolution of the number of planetesimals scattered onto Neptune
             crossing orbits for cases A, F and K (black lines) in the top, middle and 
             bottom panel respectively. Also shown is the evolution of the 
             semimajor axis of Neptune in each case, indicating the relation 
             between these 2 processes.
\label{instability_details}}
\end{figure}

The development of the dynamical instability is similar for all systems studied  
and involves a gradual increase in the number of planetesimals moving inside the disk 
inner boundary, $R_{\rm disk}$, and interacting strongly with the planets. This leads 
to a gradual increase of the orbital eccentricities of Uranus and Neptune to values such
that the MMR of these planets is broken and they can have close encounters. This is illustrated 
in Figure \ref{instability_details}, which shows the time evolution of the number of 
planetesimals in Neptune crossing orbits, i.e. planetesimals with perihelia less than 
Neptune's aphelia (at a given time), for 3 of the cases we simulated corresponding to disks
with $M_{\rm disk} = 50 M_{\rm E}$ and $R_{\rm disk} = 18$ AU. For comparison, 
we also show the behavior of Neptune's perihelia, Uranus aphelia and the eccentricity of 
their orbits.

The cumulative effect of the interaction of the dispersed planetesimals (as a result of self-gravity) 
with the ice giants leads, as the system approaches instability, to an increase in their eccentricity 
until their resonant state is broken. At such point their semi-major axes generally increase in an 
abrupt manner (as shown in Figure  
\ref{instability_details}), suggests that a threshold in the interaction between planetesimals 
and the ice giants is triggered as the number of these gradually increases. Roughly, for the 
cases shown in Figure \ref{instability_details}, the total mass that has interacted with Neptune 
at the time of the onset of the instability, $t_{\rm inst}$, is between 2 and 3 Earth masses.
This provides a possible explanation to the fact that disks with a mass
smaller than the mass of the ice giants are not capable of destabilizing the system on the 
timescales considered in this study.

\begin{figure}
\includegraphics[width=\columnwidth]{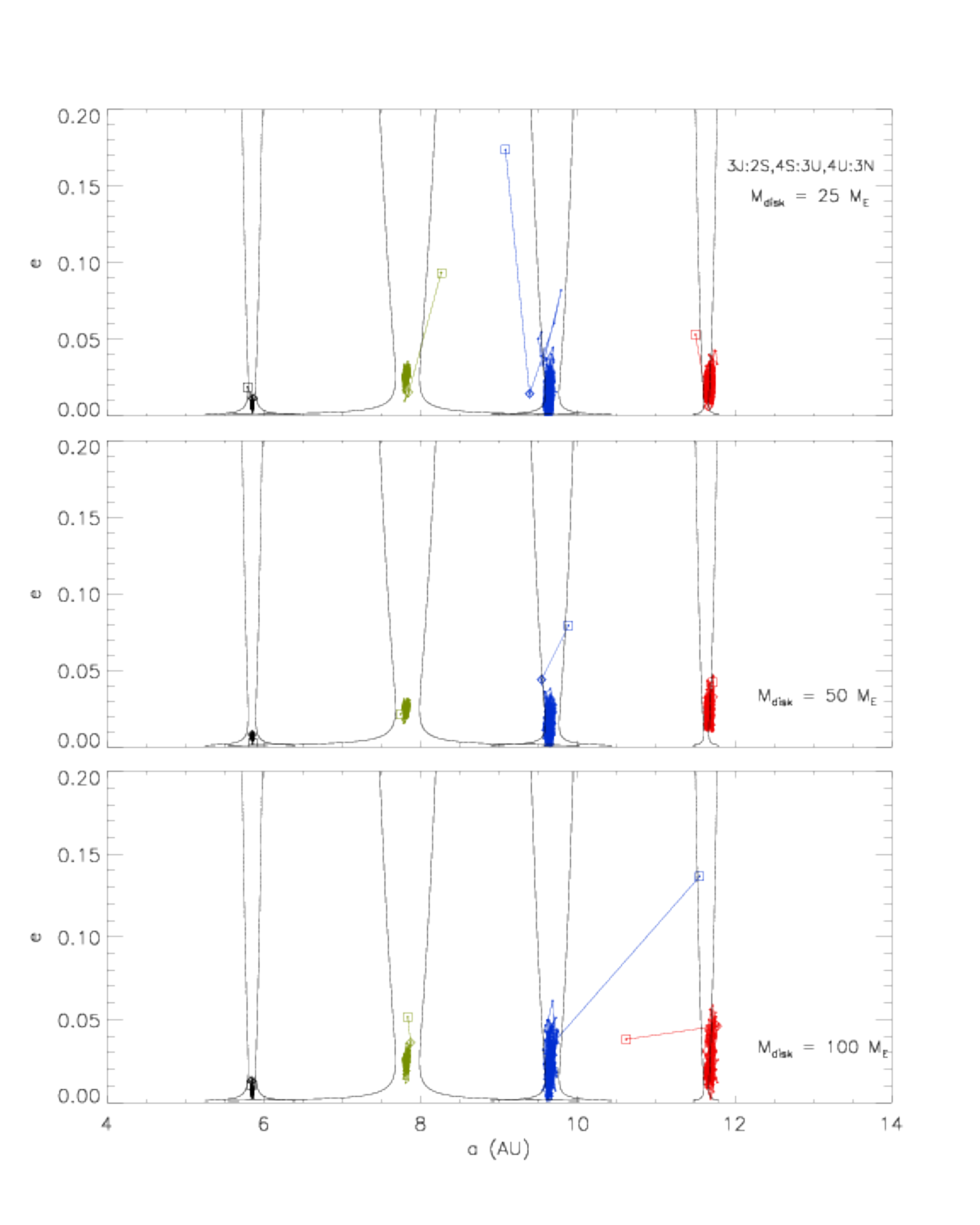}
\caption{Semi-major axis versus eccentricity diagram for each of the planets 
             as the system approaches the instability. The solid black
             lines correspond to the width of the resonance for each planet with the 
             interior planet. The diamond symbol indicates the last location of each 
             planet before the instability breaks out, the square indicates the first 
             location (in $a-e$ space) after the instability onset.
\label{instability_avse}}
\end{figure}

Figure \ref{instability_avse} shows the evolution of the planets in phase space 
($a-e$) as the system crosses the instability. The interaction with the planetesimal disk 
leads to a gradual increase in the eccentricity of the ice giants (related to angular momentum) 
while keeping the semimajor axis (energy related)  
more or less constant. Also shown are the boundaries of the resonances of Saturn, Uranus and Neptune  
with their interior planet and of Jupiter with the exterior planet. These are estimated by 
assuming that each planet is the test particle in the context of the restricted 3-body problem which, 
in the so-called pendulum approximations, allows one to estimate the width of a given first 
order resonance as described in Holmes et al. (2003). As can be seen in Figure~\ref{instability_avse}, 
the instability occurs when either Uranus or Neptune evolve toward a dynamical state in which either is 
not locked in resonance, resulting in a fast disruption of the planetary configuration.

\section{Discussion}\label{sec:discussion}

In this section we revisit some of the basic assumptions made in our study and how these 
affect the interpretation of our results. In addition, we discuss some of the implications of 
our results on the theory of planetary system evolution.

\subsection{Self-gravity}

The simulations presented above are intended to demonstrate that the inclusion of the gravitational 
interaction between the planetesimals and planets in a compact and  multiresonant initial configuration,  
as those proposed as part of the Nice model, renders the systems highly unstable. Implicit in 
this statement is the idea that, without self-gravity in the planetesimal disk, the planets in these 
systems would remain locked in resonance for a very long time, as suggested by Levison et al.~(2011).

To test the above, we carried out numerical simulations of some of the initial multiresonant conditions 
(specified in Tables 1 and 2) but removing the effect of self gravity, i.e. assuming that the planetesimals 
interact gravitationally with the planets but {\em not} among themselves (as usually done in studies
of planetesimal-driven migration). We find that some of these systems are also unstable, 
but on a timescale of several tens to hundreds of Myr. Out of the 3 systems tested in this manner 
(cases A, F and K), one is found to be unstable on a timescale of less than 100 Myr. As we are 
focusing on the effect of disk self-gravity in this study, we have decided to leave a further 
exploration of this issue for future contributions. 

Our main dynamical result brings into question the presumed stability of multiresonant configurations 
and suggests that early planetesimal driven migration, and outer disk dispersal, is to be expected for 
a wide range of initial planetary system configurations.

\subsection{Implications for the early evolution of the Solar System} 

The previous results indicate that, when gravitational interactions among 
the planetesimals are considered, several, perhaps most, of the multiresonant 
planetary configurations can become unstable on a timescale of less than a few
tens of Myr. This posses a serious difficulty for Solar System formation scenarios in 
which the LHB is explained as resulting from the unstable, fast migration of the outer 
planets and the consequent perturbation of planetesimal disk, as proposed for example 
in the Nice model.

Given the early development of the instability of multiresonant planets with a 
self-gravitating planetesimal disk, 
some of the alternative scenarios that could be envisioned for the LHB are the following.

(a) That the disk inner radius was at a distance much greater than 20 AU when the 
gaseous component of the solar nebula was dispersed, which marks the beginning 
of our simulated dynamical evolution. As can be seen from Figure~\ref{tinst_vs_Rd}, 
which shows the time for the development of the instability as a function of the inner 
disk radius of the planetesimal disk, a value of $R_{\rm disk}$ much greater than 20 AU 
is required to have an instability development time of  $t_{\rm inst} \approx 700$ Myr; that
is timescale of the LHB. 

This initial location of the inner boundary of the planetesimal disk, at such a great distance
from the outermost planet, poses in our view a difficult conundrum. While it is generally 
believed that planets as they form are able to clear material (gas and planetesimals) from an
annulus in the vicinity of their location, the so-called feeding zone (Lissauer, 1993)  is thought
to extend only to a few Hill radii from each planet's position. In the configuration suggested
above, the inner radius of the planetesimal disk is at a distance of more than 30 Hill radii  
from the outermost planet. It is not evident why planetesimals were cleared, or were not 
formed in the first place, only between approximately 12 and 25 AU.

(b) Another possible explanation to obtain a long $t_{\rm inst}$, is that the initial mass of the 
planetesimal disk was much smaller than is usually considered. Nice model studies have 
regularly used $M_{\rm disk}$ around 35  $M_{\Earth}$, and the latest incarnation of the 
model favors a slightly heavier disk of 50 $M_{\Earth}$ (Levison et al. 2011). However, we 
find that even a disk with 25 M$_{\Earth}$ can destabilize the system on a timescale more 
than an order of magnitude shorter than that required for the LHB.

\begin{figure}
\includegraphics[width=\columnwidth]{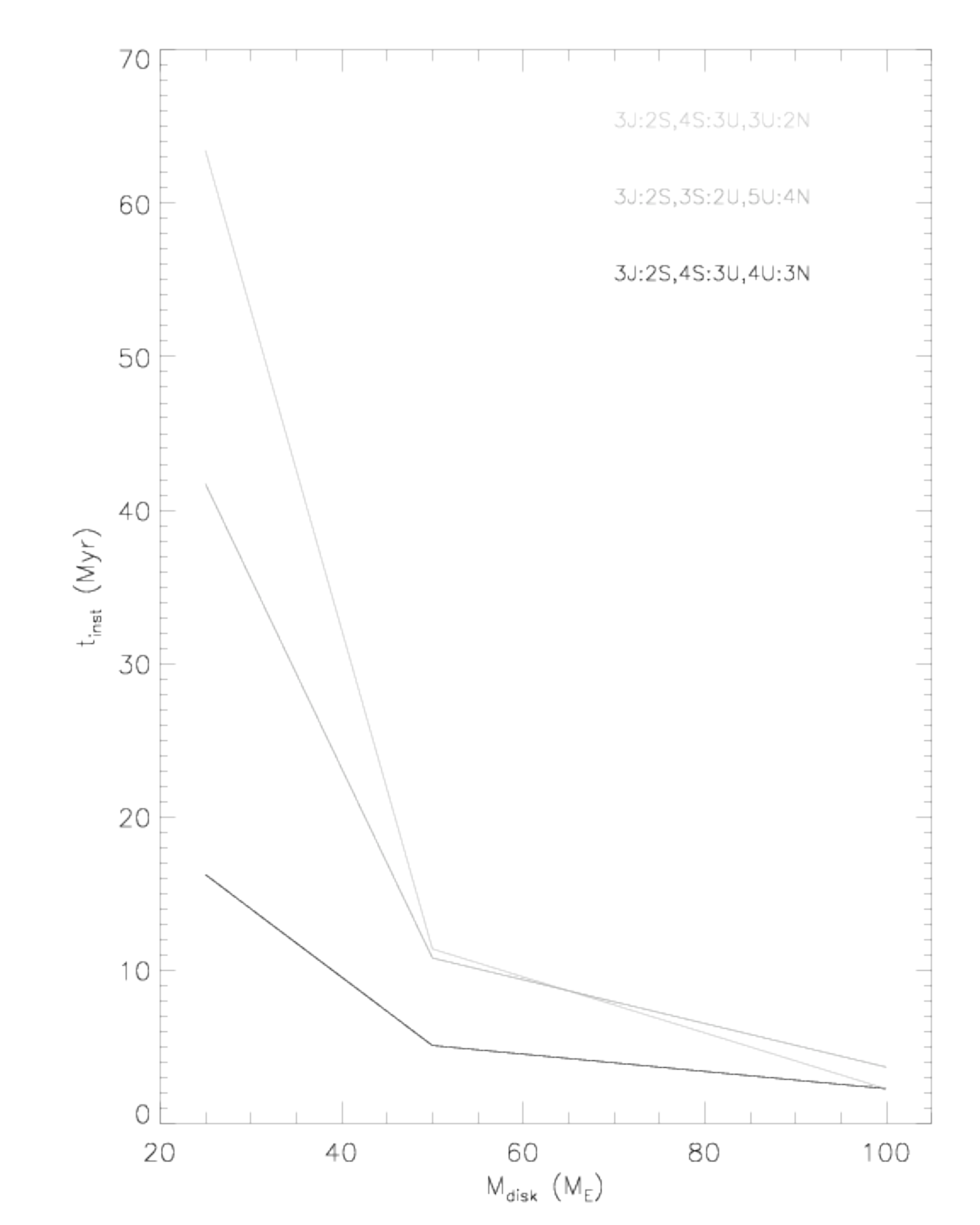}
\caption{Time at which the instability develops, $t_{\rm inst}$, as a function of initial 
             disk mass. The black line corresponds to the systems starting from the
             3:2, 4:3, 4:3 multiresonant configuration, the light gray line corresponds to
             the 3:2, 4:3, 3:2 initial configurations and the dark gray line to the systems
             starting from the 3:2, 4:3, 3:2 configuration.
\label{tinst_vs_Md}}
\end{figure}

\begin{figure}
\includegraphics[width=\columnwidth]{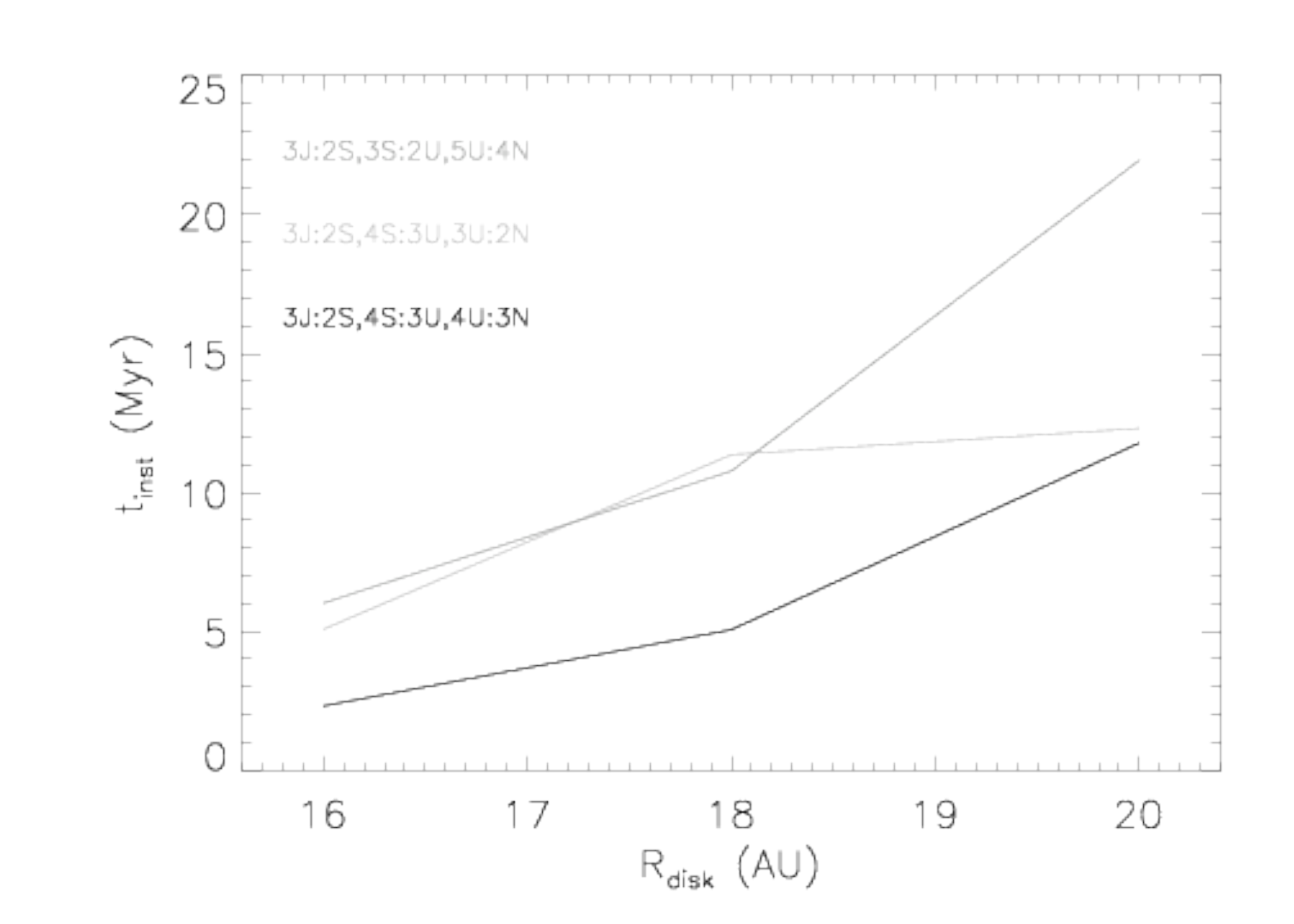}
\caption{Time at which the instability develops, $t_{\rm inst}$, as a function of initial  disk inner radius. 
The black line corresponds to the systems starting from the  3:2, 4:3, 4:3 multiresonant 
configuration, the light gray line corresponds to the 3:2, 4:3, 3:2 initial configurations and the dark 
gray line to the systems starting from the 3:2, 4:3, 3:2 configuration.
\label{tinst_vs_Rd}}
\end{figure}

Using a power law fit to the resulting timescale for the development of 
the instability as a function of the disk mass,
determined from our simulations as shown in Figure \ref{tinst_vs_Md}, suggests 
that $M_{\rm disk} \ll 20 M_{\Earth}$ is required if the LHB is to be explained 
in terms of orbital dynamical instability. However, one would expect that very low mass
disks may be altogether unable to destabilize the planetary system, as the mass
and angular momentum contained in the planetesimal disk is smaller than that of 
any of the outer planets. We can estimate a lower limit for the disk mass required 
to significantly modify Neptune's orbit by comparing the initial angular momentum of
the planet, $L_{\rm N} \approx M_{\rm N} \ V_{\rm K}(a_{\rm N}) \ a_{\rm N}$, 
with the total angular momentum in the disk, 
$L_{\rm disk} \approx M_{\rm disk} \ V_{\rm K}(R_{\rm out}) \ R_{\rm out}$. 
The requirement that $L_{\rm disk} > L_{\rm N}$, leads to the condition 
$M_{\rm disk}  \ga 10 M_{\rm E}$ for the planetesimal disk to
destabilize the outermost planet.

(c) It could also be possible that the Solar System did not start from a multiresonant, 
compact configuration as it is generally proposed in the latest incarnations of the Nice model. 
Multiresonant configurations are believed to be the natural end product of planet rearrangement 
due to gas-driven migration in the primeval gaseous solar  nebula. The presumed stability of 
multiresonant configurations is further invoked to explain the long delay for the planetary 
configuration to undergo a fast dynamical instability and the consequent stirring of the outer 
planetesimal disk and the LHB  (Levison et al. 2011). However, as we have shown in this work, 
this belief is not entirely justified. In fact, our results suggest that several compact configurations
are not stable over a 100~Myr timescale when the effects of a self-gravitating planetesimal disk, 
of structural properties similar to those assumed in the framework of the Nice model, are taken 
into account.

One could envision that the timescale necessary for the formation of all the giant 
planets and their locking into a multiresonant configuration did not exactly 
coincide with the timescale for the dispersal of the gaseous component of 
protoplanetary disks. This is particularly true if gas dispersal is related to 
photoevaporation of the disk or any other process unrelated to the formation 
of the planets themselves. In such case, the early planetary system may be left 
free of gas so that no further gas-driven migration is possible before the planets 
reach the proposed compact, multiresonant configuration.  

The subsequent evolution of these initially compact systems caught outside of 
full multiresonance at the time of gas dispersal, is beyond the scope of the present 
study. However, one would expect that if the configuration is compact enough,
the evolution could be similar to that of the original Nice model as presented in
Gomes et al. (2005) and Tsiganis et al. (2005). On the other hand, if gas driven 
migration stops while the ice giants are located far from Jupiter and Saturn, the 
evolution can be expected to proceed in a relatively smooth manner as studied by Hahn \&
Malhotra (1999) or Gomes et al. (2004). It must be stressed however, that in these studies
the effect of planetesimal-planetesimal interactions was not considered.

(d) Finally, one further possibility consistent with the present study, is that 
the unstable rearrangement of the solar system is not linked to the LHB. In 
this scenario the Solar System underwent an unstable rearrangement shortly 
after gas was dispersed, on a timescale of 10's of Myr, so that the much later 
LHB is not related to a planetesimal-driven orbital instability. Although there are strong
arguments against alternative scenarios for the origin of the LHB, such as that 
proposed by Chambers (2007) involving the effect of a small planet originally 
present between Mars and Jupiter (e.g. Brasser \& Morbidelli 2011),  it is worth 
noting that such scenario can not be entirely ruled out as a particular solution
on dynamical grounds, albeit of a low probability. This situation, is reminiscent 
of the jumping Jupiter scenario invoked in the current version of the Nice model.

\subsection{Model Limitations} 

The important implications of our study, which may 
affect our understanding of the evolution of the Solar System, hints toward the 
importance of careful modeling of the disk of planetesimals in order put on firmer 
grounds the previous conclusion.

Our numerical approach does not allow us, due to computational limitations, 
to follow the evolution  of these planetary systems in a self-consistent way  for a 
LHB-like timescale of several hundred million years. However, a simple 
extrapolation of  our instability results, as a function of disk mass and inner radius, 
suggests that if the LHB is to
be explained in terms of a dynamical instability of the early Solar System, the initial 
outer planetesimal disk must have had been  much less than 20 Earth masses and/or the 
disk inner radius 
must have been located at more than 10 AU from the outermost planet. Either condition 
has important implications for the preceding stages in the evolution of the Solar System. 
An even more  radical solution to this conundrum is the possibility that the LHB is not directly 
related to a dynamical instability of the outer planets.

Although we consider our calculations to represent the state of the art in the 
study of planetesimal-driven migration, on account of their self-consistent treatment, 
it is worth pointing out some factors related to our present computational limitations that 
may affect our conclusions. In our study each particle has a mass  
ranging from (0.75--3)$\times 10^{26}$g that is $\approx (4-17)$ times the mass 
of Eris, the largest of the dwarf planets. Considering that the present mass of the Kuiper Belt 
is believed to be $\approx 1/500$ of the mass assumed  to be originally present in the 
outer planetesimal disk (Gladman et al. 2001), and assuming that the size distribution of 
large planetesimals was originally similar to the present one in the KB, one would expect 
that the original number of dwarf planet-like bodies to be in the vicinity of a few times 
10$^3$. This implies that each of the small bodies in our simulations corresponds to 
approximately 5 dwarf planet-like bodies originally believed to be present in the Solar System. 

The above approximation does not seem too bad, but it is known that the number of particles, 
$N$, in an $N$-body simulation affects the dynamical evolution of self-gravitating systems  
(e.g. Binney \& Tremaine 1998). It is possible that an order of magnitude increase in the 
number of small bodies, as predicted by the present day size distribution of KBOs, can have 
an important effect on the dynamics of the Solar System. For example, on one hand, 
the presence of a very large number of small bodies allows for a further distribution of the 
interaction-energy among these by incrementing the number of degrees of freedom of the 
whole system. In this way, the small bodies may behave as a stabilizing sink, that is 
probably dependent on their number, for the development of the instability as is suggested 
by the results of this paper. On the other hand, the actual behavior of the disk particles is 
dependent on the number of particles used to describe it in what is usually called numerical 
disk-heating. In this process the velocity dispersion of the disk particles, $\sigma_{\rm p}$, 
modeled as a diffusion process, grows with time $t$ as 
$\propto (\sigma_0^2 + D_{\rm p} t)^p$; where $D_{\rm p}$ is a diffusion coefficient, 
$p$ a parameter and $\sigma_0$ a constant (e.g. Lacey 1991, Vel\'azquez~2005). The 
values of $D_{\rm p}$ and $p$ are model dependent and they  have to be evaluated 
numerically under specific physical conditions. However they show a dependency on the 
number of particles in the system, for example, $D_{\rm p} \propto N^{-\alpha}$, with 
$\alpha \approx 0.6$ for galactic disks (Vel\'azquez~2005). Nonetheless, they point to 
the fact of an effect of the number of particles used in the modeling for the evolution 
of the planetesimal disk. All this lead us to suggest that future works should consider 
some of these matters, although at present they currently represent a formidable 
computational challenge.

\section{Conclusions}\label{sec:conclusions}

We have carried out a series of direct numerical simulations of the  
the dynamical evolution of the early outer Solar System in the context of the Nice-2 model. 
Our aim has been to study this process, for the first time, in a fully self-consistent manner, i.e. 
by taking into account the gravitational interaction between all bodies included in the 
simulations. 

Our starting point was a set of initially compact, multiresonant planetary configurations, similar 
to those proposed in the study of Levison et al. (2011).
In difference with previous attempts to model such systems in the framework of
an approximate treatment of the N-body problem (e.g. Levison et al.~2011), we find that a 
dynamical instability of the outer planets develops on a short timescale, of less than 70 Myr 
in all the initial configurations we explored. The early development of the instability is directly 
related to the effect of planetesimal-planetesimal interactions, as the resulting dynamical heating 
of the planetesimal disk accelerates the interaction with the outer ice giants which leads to 
the planetary instability.

The above result has important consequences on our understanding of the dynamics of the
early Solar System. Our results suggest that when the inter-planetesimal gravitational interactions 
are included in a fully self-consistent manner, the dynamical instability of the early Solar System 
invoked in the context of the Nice model to explain the LHB, occurs much sooner than the several 
hundred Myr timescale indicated by the lunar crater record (Tera et al. 1974). Although further 
studies are required, particularly increasing the number of particles representing the planetesimal disk,
these results bring into question the explanation of the LHB in terms of a planetary dynamical instability.

\acknowledgments
The authors are grateful to Dr.~Konstantin Batygin for kindly providing the multiresonant initial 
conditions used in this study, and to Prof. Renu Malhotra for valuable comments on a draft of this paper. 
MRR acknowledges financial support from DGAPA-UNAM project IN115429, HA from CONACyT 
project 179662, and all the authors from DGAPA-UNAM project IN108914.

.  




\clearpage

\clearpage


\begin{table}
\begin{center}
\caption{Planet configuration and disk parameters of each of the cases studied\label{tbl-1}}
\begin{tabular}{ccrr}
\tableline\tableline
Case & Planet Configuration & $M_{\rm disk}$ & $R_{\rm disk}$ \\
  & J:S , S:U, U:N & M$_{\Earth}$ & R$_{\Earth}$ \\
\tableline
A & 3:2, 4:3, 4:3 & 50 & 18 \\
B & 3:2, 4:3, 4:3 & 25 & 18 \\
C & 3:2, 4:3, 4:3 & 100 & 18 \\
D & 3:2, 4:3, 4:3 & 50 & 16 \\
E & 3:2, 4:3, 4:3 & 50 & 20 \\
F & 3:2, 4:3, 3:2 & 50 & 18 \\
G & 3:2, 4:3, 3:2 & 25 & 18 \\
H & 3:2, 4:3, 3:2 & 100 & 18 \\
I & 3:2, 4:3, 3:2 & 50 & 16 \\
J & 3:2, 4:3, 3:2 & 50 & 20 \\
K & 3:2, 3:2, 5:4 & 50 & 18 \\
L & 3:2, 3:2, 5:4 & 25 & 18 \\
M & 3:2, 3:2, 5:4 & 100 & 18 \\
N & 3:2, 3:2, 5:4 & 50 & 16 \\
O & 3:2, 3:2, 5:4 & 50 & 20 \\
\tableline
\end{tabular}
\end{center}
\end{table}

\clearpage

%
\begin{table}
\begin{center}
\caption{Initial value of the main orbital elements for the planets in the multiresonant, 
             initial configurations we considered \label{tbl-2}}
\begin{tabular}{lccc}
\tableline\tableline
Planet & $a$ & $e$ & $i$  \\
       & AU  &     &  deg   \\  
\tableline\tableline
3J:2S,4S:3U,4U:3N \\
\tableline
J    &  5.84724  &  0.00690581  &  0  \\  
S    &  7.83006  &  0.0260594  &  0   \\  
U    &  9.67303  &  0.0163112  &  0     \\
N    &  11.6361  &   0.0179751  &  0    \\
\tableline\tableline
3J:2S,4S:3U,3U:2N \\
\tableline
J    &  5.87084   &  0.0037631  &  0  \\  
S    &  7.99698   &  0.0165349  &  0   \\  
U    &  9.98186   &  0.0168401  &  0     \\
N    &  13.1645   &   0.0064222  &  0    \\
\tableline\tableline
3J:2S,3S:2U,5U:4N \\
\tableline
J    &  5.88018   &  0.00597505  &  0  \\  
S    &  7.88684   &  0.0245716  &  0   \\  
U    &  10.3836   &  0.0305705  &  0     \\
N    &  12.005   &   0.00827154  &  0    \\
\tableline\tableline
3J:2S,3S:2U,4U:3N (Levison et al. 2011) \\
\tableline
J    &  5.42   &  0.0044  &  0.016  \\  
S    &  7.32   &  0.017  &  0.016   \\  
U\tablenotemark{a} &  9.61   &  0.053  &  0.044     \\
N\tablenotemark{a}    &  11.67   &   0.011  &  0.029    \\
\tableline
\end{tabular}
\tablenotetext{a}{ Levison et al. (2011) do not consider Uranus and Neptune specifically,
                          but {\it generic} 15 $M_{\rm E}$ ice giants instead. Indicated values correspond 
                          to reported average values for these parameters.}
\end{center}
\end{table}
%


\end{document}